\documentclass[
superscriptaddress,
 amsmath,amssymb,
 aps, physrev,
 twocolumn,
]{revtex4-2}

\usepackage{graphicx}
\usepackage{dcolumn}
\usepackage{bm}
\usepackage{hyperref}
\usepackage{xcolor}

\begin{document}


\title{\textbf{An effective finite-range Gogny-type interaction for the quantum molecular dynamics like model} 
}%
 
\author{Meiqi Sun}
 \affiliation{China Institute of Atomic Energy P. O. Box 275(18) Beijing 102413 China}

 \author{Dandan Niu}
 \affiliation{China Institute of Atomic Energy P. O. Box 275(18) Beijing 102413 China}

\author{Junping Yang}
 \affiliation{%
College of Physics and Optoelectronic Engineering Shenzhen University Shenzhen 518060 China
}%
 
\author{Ying Cui}
 \affiliation{China Institute of Atomic Energy P. O. Box 275(18) Beijing 102413 China}

\author{Zhuxia Li}
 \affiliation{China Institute of Atomic Energy P. O. Box 275(18) Beijing 102413 China}

\author{Qiang Zhao}
\affiliation{China Institute of Atomic Energy P. O. Box 275(18) Beijing 102413 China}

\author{Kai Zhao}
\email{Contact author: zhaokai@ciae.ac.cn}
 \affiliation{China Institute of Atomic Energy P. O. Box 275(18) Beijing 102413 China}

\author{Yingxun Zhang}%
 \email{Contact author: zhyx@ciae.ac.cn}
 \affiliation{China Institute of Atomic Energy P. O. Box 275(18) Beijing 102413 China}
\affiliation{Department of Physics and Technology Guangxi normal University Guilin 540101 China}

\date{\today}

\begin{abstract}
In this work, we propose an effective finite-range Gogny-type interaction that can be directly used in the quantum molecular dynamics (QMD) like model. Two methods for determining the parameters of the effective interaction are discussed. The first method establishes an approach to connect the conventional Gogny interaction in nuclear structure to that in heavy-ion collisions, the second method allows for the description of the symmetry energy varying from the supersoft to stiff, as well as the momentum-dependent symmetry potential, exhibiting behaviors ranging from monotonic to non-monotonic variations. This effective interaction opens up opportunities for a deeper understanding of finite-range interactions and non-monotonic momentum-dependent symmetry potentials in future studies.

\end{abstract}

\maketitle



The nature of the effective nucleon-nucleon interaction is central to understanding the properties of complex nuclei and the mechanism of heavy ion reactions or collisions. Historically, the effective nucleon-nucleon interaction was built in with two primary approaches: the zero-range effective interaction~\cite{Skyrme1956PM} and the finite-range effective interaction~\cite{Gogny1980PRC}. 
Both types of effective interaction were widely used in theoretical models to study the properties of nuclear structure and its related issues. For example, the zero-range Skyrme effective nucleon-nucleon interaction, parametrized by its dependence on the relative distance and its derivatives, has been extensively used in studies of nuclear structure\cite{Vautherin1972PRC,YZWang2020CTP}, mass\cite{Kohler1976NPA,Goriely2002PRC,Pearson2006NPA,Aichelin2020PRC}, energy level\cite{Sagawa1985PLB,Hilaire2006NPA,Valuev2020PRA}, fission\cite{Dutta1980NPA,Schunck2014PRC,Sharma2017EPJA} and properties of neutron stars~\cite{Chabanat1997NPA,Rikovska2003PRC,Erler2013PRC,Nandi2016PRC,Mehta2022Universe}, owing to its ease of implementation in practical calculations. The finite-range interaction is generally considered more physically realistic 
and has also been employed in the studies of nuclear fission\cite{BERGER1991CPC} and deformation\cite{Robledo2011PRC}, multipolar collective degrees of freedom\cite{Gaffney2013Nature,Peru2014EPJA}, and in the studies of neutron stars\cite{Roshan2014PRC,Gonzalez2017PRC,Gonzalez2018PLB,Mondal2020PRC,Vinas2021Symmetry}.

For heavy ion collisions, the finite-range interaction becomes more important and indispensable. For example, the exchange term on the finite-range interaction generates a momentum-dependent potential, which plays a crucial role in describing the collective flow with a reasonable nuclear matter incompressibility parameter\cite{Aichelin1987PRL}. In the QMD-like models and BUU (Boltzmann-Uhling-Uhlenbeck)-like models, one usually adopted either the zero-range Skyrme interaction\cite{YXZHANG2014PLB,YXZhang2020FOP} or a simple zero-range interaction plus a phenomenological momentum dependent interaction\cite{Aichelin1985PRL,QFLi2011PRC,HARTNACK1989NPA,Cozma2013PRC,YJWang2020FOP} to study the nuclear equation of state and the properties of nuclear medium. Up to now, there are some efforts to use the finite-range interaction in the transport models. For instance, the isospin Boltzmann-Uhling-Uhlenbeck (IBUU) transport model\cite{Das2003PRC,BaoanLi2005PRL,JunXU2011PRC,LWChen2012PRC,HuiXue2016PLB}, the Antisymmetrized Molecular Dynamics (AMD) model\cite{Akira2003}, the constrainted molecular dynamics (CoMD) model~\cite{PAPA2024NPA} and dcQMD~\cite{Cozma2021EPJA} have employed the Gogny-like interactions. But one should note that these interactions are specific for heavy ion collisions, and have not been directly related to the conventional Gogny interaction yet. 



In addition, the constraints on the isospin asymmetric nuclear equation of state (EOS) with the transport models depend on the exact form of the momentum-dependent symmetry potential\cite{BALi2004NPA,Rizzo2005PRC,Rizzo2010PRC,YXZHANG2014PLB}. Especially, the recent reanalysis of the neutron to proton yield ratio by Yang \textit{et al.}\cite{JPYANG2024PRC} suggests that the momentum-dependent symmetry potential may first decrease and then increase, which cannot be explained by using the published 246 sets of the zero-range Skyrme interaction\cite{Dutra2012PRC}. Luckily, the studies in Refs.~\cite{LWChen2012PRC,Roshan2014PRC} have found that the Gogny D1S\cite{BERGER1991CPC} and D250\cite{BLAIZOT1995NPA} can predict that the momentum-dependent symmetry potential first decreases and then increases. Thus, one may expect to find or built a Gogny-type interaction, which can be used in the QMD model and may answer the above question.


The present work aims to derive a parametrized version of the Gogny-type interaction, that enables broad adjustability of the symmetry energy stiffness and can be directly related to the conventional Gogny interactio. 
The paper is organized as follows: first, we will briefly review the framework of the quantum molecular dynamics approach and then mention the proposed finite-range Gogny interaction. Then, we will discuss the corresponding equation of state, symmetry energy, single particle potential and symmetry potential, and the effective mass of neutron and proton. Finally, we will give the results on the potential energy within the framework of QMD-like model for practical utilization.


In the QMD approach\cite{AICHELIN1991PR,YXZhang2020FOP}, each nucleon is represented by a Gaussian wave packet,\vspace*{0.5mm}
\begin{equation}
\label{eq:singlewf}
\phi_i(\mathbf{r}_i)=\frac{1}{(2\pi\sigma_r^2)^{3/4}}{\rm e}^{-\frac{(\mathbf{r}_i-\mathbf{r}_{i0})^2}{2\sigma_r^2}+{\rm i}(\mathbf{r}_i-\mathbf{r}_{i0})\cdot \mathbf{p}_{i0}/\hbar},\vspace*{0.5mm}
\end{equation}
here, $\sigma_r$ and $\mathbf{r}_{i0}$ are the width and centroid of
wave packet, respectively. The subscript $i$ of singe particle wave function $\phi_i$ represents the state that is described by position $\mathbf{r}_{i0}$, momentum $\mathbf{p}_{i0}$, and isospin $\tau_i$. One should note that the effect of the spin on the Fermi momentum of particle $i$ is considered in the QMD-like model, even the spin wave function do not appear in above formula. The system wave function $\Psi$ is assumed as a direct product of $N$ single particle wave functions, i.e., in the Hartree approximation,\vspace*{0.5mm}
\begin{eqnarray}
\label{eq:syswf}
\Psi(\mathbf{r}_1, \cdots  ,\mathbf{r}_N)=\phi_{1}(\mathbf{r}_1)\phi_{2}(\mathbf{r}_2)\cdots  \phi_{N}(\mathbf{r}_N).
\end{eqnarray}
Correspondingly, the energy of system is calculated as,
\begin{equation}
 E=\langle \Psi |\mathrm{\hat{H}}|\Psi\rangle,    
\end{equation}
or calculated with the Wigner density function as in Ref.\cite{AICHELIN1991PR}. If we only consider the effective two-body interaction $v_{ij}$, the Hamiltonian can be written as,
\begin{equation}
    \hat{\mathrm{H}}=\sum_{i}\frac{\hat{\mathbf{p}}_i^2}{2m}+\sum_{i<j} \hat{v}_{ij}.
\end{equation}
Here, $\hat{\mathbf{p}}_i$ is the momentum of $i$th nucleon, $m$ is the nucleon mass, and $\hat{v}_{ij}$ is the effective interaction used in the QMD-like models. 

Within the framework of the QMD-like models, we propose an effective finite-range Gogny-type interaction as follows,
\begin{equation}
    \label{eq:interaction}
    \begin{aligned}
 		\hat{v}_{ij}&=\sum_{l=1}^{2} \bigg[\left( \tilde{A}_l+\tilde{B}_l\hat{P}_\tau\right)e^{-\frac{(\mathbf{r}_1-\mathbf{r}_2)^2}{\mu_l^2}}\\
        &\quad +\left( \tilde{C}_l+\tilde{D}_l\hat{P}_\tau\right)e^{-\frac{\mu_l^2(\mathbf{k}_i-\mathbf{k}_j)^2}{4}}e^{-\frac{(\mathbf{r}_1-\mathbf{r}_2)^2}{\mu_l^2}}\bigg]\\
        &\quad+\left(\beta_0+\beta_1\hat{P}_\tau\right)\rho^{\sigma}(\frac{\mathbf{r}_1+\mathbf{r}_2}{2})\delta(\mathbf{r}_1-\mathbf{r}_2).\\
 	\end{aligned}
\end{equation}
It consists of two finite-range two-body terms and a density-dependent zero-range two-body term. The density-dependent zero-range two-body term provides a phenomenological representation of many-body effects\cite{Vautherin1972PRC}. The operators $\hat{P}_\tau$ is the isospin exchange operators. The terms related to $\tilde{A}_l$ and $\tilde{B}_l$ come from the direct term, and the terms related to $\tilde{C}_l$ and $\tilde{D}_l$ are used to mimic the exchange term which results in the momentum-dependent potential. The terms $\beta_0$ and $\beta_1$ have similar meanings but for density-dependent zero-range interaction. The parameters $\mu_1=0.7$ fm and $\mu_2=1.2$ fm, which are the same as those in the Gogny interaction in Refs.\cite{Gogny1980PRC}. For convenience, we named the proposed interaction used in the QMD-like models as QG interaction in the following discussions.

In this work, we take two ways to determine the 11 parameters in Eq.(\ref{eq:interaction}), i.e., $\tilde{A}_{l=1,2}$, $\tilde{B}_{l=1,2}$, $\tilde{C}_{l=1,2}$, $\tilde{D}_{l=1,2}$, $\beta_0$, $\beta_1$ and $\sigma$, by describing the EOS of nuclear matter reasonably. 

The first way is to determine the parameters from the conventional Gogny interactions\cite{Gogny1980PRC}. As known in Ref.\cite{Gogny1980PRC}, the conventional Gogny interaction $\hat{v}_{ij}^*$ taken as,
\begin{equation}\label{eq:gogny}
    \begin{aligned}
        v_{ij}^*&=\sum_{k=1}^{2}\Bigg( W_k+B_k\hat{P}_\sigma-H_k\hat{P}_\tau-M_k\hat{P}_\sigma \hat{P}_\tau\Bigg) e^{-(\mathbf{r}_1-\mathbf{r}_2)^2/\mu_k^2}\\
        &\quad\quad+t_0(1+x_0 \hat{P}_\sigma)\rho^{\sigma} (\frac{\mathbf{r}_1+\mathbf{r}_2}{2})\delta(\mathbf{r}_1-\mathbf{r}_2).\\
    \end{aligned}
\end{equation}
The $W_k$, $B_k$, $H_k$, and $M_k$ are the parameters in the conventional Gogny interactions. The idea for the determination of parameters is that the potential energy obtained within the Hartree-Fock approximation is equal to the potential energy obtained within the Hartree approximation in QMD approach, i.e., 
\begin{equation}\label{eq:relation-gogny-qmd}
    \begin{aligned}
        &\sum_{ij\sigma\sigma'\tau\tau'}\langle ij\sigma\sigma'\tau\tau'|\hat{v}_{ij}^*(1-\hat{P}_{i\leftrightarrow j})|ij\sigma\sigma'\tau\tau'\rangle\\
        &=4\sum_{ij\tau\tau'}\langle ij\tau\tau'|\hat{v}_{ij}|ij\tau\tau'\rangle.
    \end{aligned}
\end{equation}
The $\hat{P}_{i\leftrightarrow j}$ is the exchange operator, and is equal to the product of Majorana exchange $\hat{P}_M$, spin exchange $\hat{P}_\sigma$, and isospin exchange operator $\hat{P}_\tau$, i.e., $\hat{P}_M \hat{P}_\sigma \hat{P}_\tau$. The prefactor $4$ in the right hand of Eq.(\ref{eq:relation-gogny-qmd}) is from the summation of spin of particle $i$ and $j$, since there is no explicit spin degree in the QMD-like model. 

\begin{figure}
    \centering
    \includegraphics[width=1.0\linewidth]{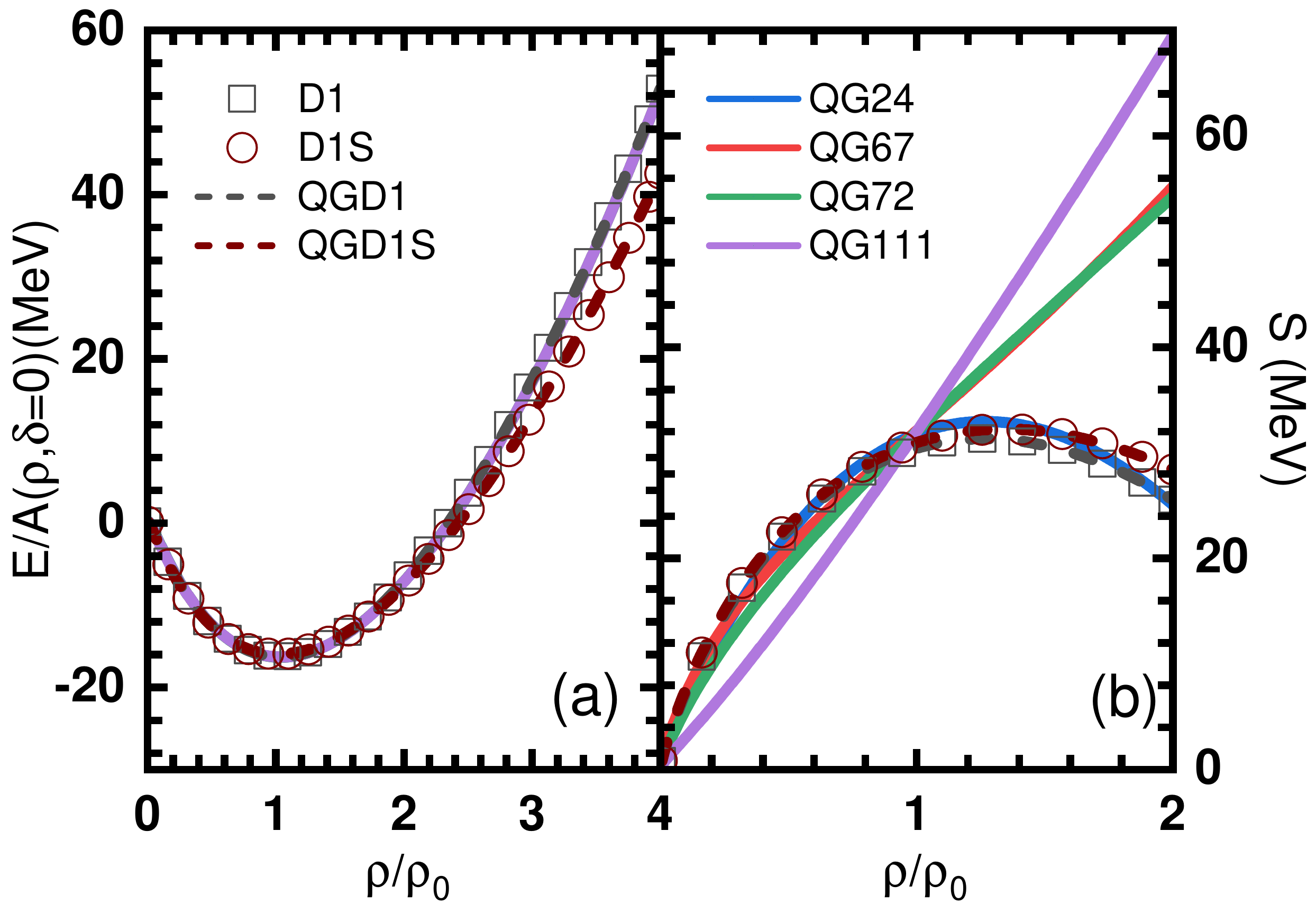}
    \caption{(Color online) The isospin symmetric nuclear equation of state (a), and the density dependence of the symmetry energy (b) obtained with different QG interactions (lines). The symbols are results obtained with the conventional Gogny interaction. }
    \label{fig:EoS&SYMMETRY_ENERGY}
\end{figure}

Based on Eq.(\ref{eq:relation-gogny-qmd}), the following relationship can be obtained,
\begin{equation}
        \begin{aligned}
        &\tilde{A}_l=W_k+\frac{1}{2}B_k,\\
        &\tilde{B}_l=-(H_k+\frac{1}{2}M_k),\\
        &\tilde{C}_l=\frac{1}{2}H_k+M_k,\\
        &\tilde{D}_l=-(\frac{1}{2}W_k+B_k),\\
        &\beta_0=t_0(1+\frac{1}{2}x_0),\\
        &\beta_1=-t_0(\frac{1}{2}+x_0).
    \end{aligned}
\end{equation}
Since the parameters are determined by the conventional Gogny parameter sets D1 and D1S, we named them as QGD1 and QGD1S. One can expect that the EOS and symmetry energy will be identical for QG interaction sets (QGD1, QGD1S) and the conventional Gogny interaction sets (D1, D1S) as in Figure~\ref{fig:EoS&SYMMETRY_ENERGY} (a) and (b). 

The second way is to directly fit symmetric nuclear matter EOS with the effective QG interaction $\hat{v}_{ij}$. In this case, the parameters $\tilde{A}_{l=1,2}$, $\tilde{B}_{l=1,2}$, $\tilde{C}_{l=1,2}$, $\tilde{D}_{l=1,2}$, $\beta_0$, $\beta_1$ and $\sigma$ are not solely determined by the known symmetric EOS, and thus the different stiffness of symmetry energy can be obtained. In this work, we fit the symmeric EOS obtained with Gogny D1 set within 3 times the normal density in the case of the reduced $\chi^2<1$. 

Now, let's check the nuclear equation of state, single-particle potential, and nuclear matter parameters in cold nuclear matter for QG interaction. To describe the uniform nuclear matter within the framework of QMD wave function, we need to set $\sigma_r\to \infty$ and thus the Gaussian wave packet for each nucleon in Eq.(\ref{eq:singlewf}) tend to the plane wave, 
\begin{equation}
\begin{aligned}
    &\phi_i=\frac{1}{\sqrt{V}}e^{i\mathbf{k}\cdot\mathbf{r}}. 
\end{aligned}
\end{equation}
Consequently, the expression of isospin asymmetric nuclear EOS for cold nuclear matter is,
\begin{equation}\label{eq:asyeos}
    \begin{aligned}
        \frac{E}{A}(\rho,\delta)&=\sum_{i\sigma\tau}\langle i\sigma\tau |\frac{\hat{p}_i^2}{2m}|i\sigma\tau\rangle\\
        &\quad\quad +\frac{1}{2}\sum_{ij\sigma\sigma'\tau\tau'}\langle ij\sigma\sigma'\tau\tau' |\hat{v}_{ij}|ij\sigma\sigma'\tau\tau'\rangle\\
        &=\frac{3\hbar^2}{20m}\left(\frac{3\pi^2}{2}\rho\right)^{\frac{2}{3}}[(1+\delta)^{5/3}+(1-\delta)^{5/3}]\\
        &\quad+\sum_{l=1}^{2}\frac{(2\tilde{A}_l+\tilde{B}_l)}{4}(\sqrt{\pi}\mu_l)^3\rho+\sum_{l=1}^{2}\frac{\tilde{B}_l}{4}(\sqrt{\pi}\mu_l)^3\delta^2\rho\\
        &\quad+\sum_{l=1}^{2}\frac{2(\tilde{C}_l+\tilde{D}_l)}{3\mu_l^3\pi^{\frac{5}{2}}\rho} \times\left[{e}(\mu_l k_{fn})+{e}(\mu_l k_{fp})\right]\\
        &\quad+\sum_{l=1}^{2} \frac{2\tilde{C}_l}{3\mu_l^3\pi^{\frac{5}{2}}\rho} \times{\bar{e}}(\mu_l k_{fn},\mu_l k_{fp})\\
        &\quad+\frac{1}{4}(2\beta_0+\beta_1)\rho^{\sigma+1}+\frac{1}{4}\beta_1\rho^{\sigma+1}\delta^2.
    \end{aligned}
\end{equation}
Here, $\rho=\rho_n+\rho_p$ with the neutron density $\rho_n$ and $\rho_p$. The $\delta=(\rho_n-\rho_p)/(\rho_n+\rho_p)$ is the isospin asymmetry. 
One should keep in mind that the system wave function is the direct product of single-particle wave function, and thus the formula of $E/A$ in Eq.(\ref{eq:asyeos}) is different than the results in Ref.\cite{Roshan2014PRC,LWChen2012PRC}. The function ${e}(\eta)$ and ${\bar{e}}(\eta_n,\eta_p)$ have the following expressions,
    \begin{equation}
    \begin{aligned}
    &{e}(\eta)=\frac{\sqrt{\pi}}{2}\eta^3\mathrm{erf}(\eta)+(\frac{\eta^2}{2}-1)e^{-\eta^2}-\frac{3\eta^2}{2}+1\\
    &{\bar{e}}(\eta_n,\eta_p) = \frac{\sqrt{\pi}}{2}\bigg[(\eta_n^3+\eta_p^3)\mathrm{erf}\left(\frac{\eta_n+\eta_p}{2}\right)\\
    &\quad-(\eta_n^3-\eta_p^3)\mathrm{erf}\left(\frac{\eta_n-\eta_p}{2}\right)\bigg]\\
    & \quad + (\eta_p^2-\eta_n\eta_p+\eta_n^2-2)\exp\bigg[-\frac{(\eta_n+\eta_p)^2}{4}\bigg]\\
    &\quad- (\eta_p^2+\eta_n\eta_p+\eta_n^2-2)\exp\bigg[-\frac{(\eta_n-\eta_p)^2}{4}\bigg].
    \end{aligned}
\end{equation}
Here, the $erf()$ is the error function.

According to the Eq.(\ref{eq:asyeos}) and the definition of the nuclear matter parameters, we can get the incompressibility $K_0$, i.e.,
\begin{equation}
\begin{aligned}
    K_0
    & = 9\rho^2\frac{\partial^2(E_0/A)}{\partial\rho^2}|_{\rho_0}\\
    & = -\frac{3\hbar^2}{5m}\left(\frac{3\pi^2}{2}\right)^{\frac{2}{3}}\rho_0^{\frac{2}{3}}+\frac{9\sigma(\sigma+1)}{4}(2\beta_0+\beta_1)\rho_0^{\sigma+1}\\
    &\quad+6\sum_{l=1}^2\frac{2\tilde{C}_l+\tilde{D}_l}{\sqrt{\pi}}\Bigg[\frac{6}{(\mu_l k_f)^3}-\frac{2}{\mu_l k_f}\\
    &\quad\quad\quad -e^{-(\mu_l k_f)^2}\left(\frac{6}{(\mu_l k_f)^3}+\frac{4}{\mu_l k_f}+\mu_l k_f\right)\Bigg].\\
\end{aligned}
\end{equation}
Here, $k_f=(\frac{3\pi^2}{2}\rho_0)^{1/3}$.



Furthermore, we can obtain the density dependence of the symmetry energy by derivativing the Eq.(\ref{eq:asyeos}) over isospin asymmetry $\delta$. The expression reads
\begin{equation}
    \begin{split}
        S(\rho)
        & = \frac{1}{2!}\frac{\partial^2 E(\rho,\delta)}{\partial\delta^2}\bigg|_{\delta=0}\\
        & = \frac{\hbar^2}{6m}\left(\frac{3\pi^2}{2}\rho\right)^{\frac{2}{3}}+\frac{1}{4}\sum_{l=1}^{2} \tilde{B}_l(\sqrt{\pi}\mu_l)^3\rho\\
        &\quad+\sum_{l=1}^{2}\frac{\tilde{C}_l}{3\sqrt{\pi}}\left[{S}_1(\mu_lk_f)+{S}_2(\mu_lk_f)\right]\\
        &\quad+\sum_{l=1}^{2}\frac{\tilde{D}_l}{3\sqrt{\pi}}{S}_1(\mu_lk_f)+\frac{1}{4}\beta_1\rho^{\sigma+1},
    \end{split}
\end{equation}
with
\begin{align}
     &{S}_1(\eta)=\frac{1}{\eta}-\left(\eta+\frac{1}{\eta}\right)e^{-\eta^2},\\
     &{S}_2(\eta)=\frac{1}{\eta}-\left( \eta+\frac{e^{-\eta^2}}{\eta} \right).
\end{align}
Correspondingly, the slope of symmetry energy at normal density can be written as,
\begin{equation}
    \begin{aligned}
        L
        & = 3\rho_0\frac{\partial S(\rho)}{\partial\rho}\bigg|_{\rho_0}\\
        & = \frac{\hbar^2}{3m}\left(\frac{3\pi^2}{2}\right)^{\frac{2}{3}}\rho_0^{\frac{2}{3}}+\frac{3}{4}\sum_{l=1}^2\tilde{B}_l(\sqrt{\pi}\mu_l)^3\rho_0\\
        &\quad+\frac{3(\sigma+1)}{4}\beta_1\rho_0^{\sigma+1}\\
        &\quad+ \sum_{l=1}^2\frac{\tilde{C}_l+\tilde{D}_l}{3\sqrt{\pi}} {L}_1(\mu_l k_f) + \sum_{l=1}^2\frac{\tilde{C}_l}{3\sqrt{\pi}} {L}_2(\mu_l k_f)\\
    \end{aligned}
\end{equation}
with, 
\begin{align}
    &{L}_1(\eta) = -\frac{1}{\eta} + e^{-\eta^2} \left(\frac{1}{\eta} + 2\eta^3 + \eta\right)\\
    &{L}_2(\eta) = -\frac{1}{\eta} - \eta + 2\eta e^{-\eta^2} + \frac{e^{-\eta^2}}{\eta}.
\end{align}


In Figure \ref{fig:EoS&SYMMETRY_ENERGY} (a), we present the EOS obtained from different QG interactions, i.e., QG24, QG67, QG72, and QG111. These sets are obtained by fitting the isospin symmetric EOS obtained by Gogny D1 interaction\cite{Gogny1980PRC}. The number following the `QG' refers to the value of the slope of symmetry energy, for example, QG67 means the sets has the slope of symmetry energy at normal density $L$ is 64 MeV. Since we only fit the isospin symmetric EOS, the values of parameters $\tilde{A}_l$, $\tilde{B}_l$, $\tilde{C}_l$, $\tilde{D}_l$, $\beta_0$, $\beta_1$ and $\sigma$ are not identical to the conventional Gogny D1 interaction according to Eq.(\ref{eq:relation-gogny-qmd}). There are multiple-solutions for the parameters $\tilde{A}_l$, $\tilde{B}_l$, $\tilde{C}_l$, $\tilde{D}_l$, $\beta_0$, $\beta_1$ and $\sigma$ for fitting the isospin symmetric EOS, and result in different stiffness of the symmetry energy as in Figure \ref{fig:EoS&SYMMETRY_ENERGY} (b). For the convenience of the utility of QG interaction in the framework of QMD for studying the density-dependent of symmetry energy and momentum dependent symmetry potential, we also provide 15 sets of QG interaction in Appendix~\ref{app:15qg} Table~\ref{tab:parameters}. 

%

The single-particle potential $U_\tau$ in cold nuclear matter is obtained with a similar method as in Ref.\cite{Roshan2014PRC},
\begin{equation}
    \begin{aligned}
        U_{\tau}
        & =\frac{1}{2}\sum_j\sum_{\sigma,\sigma^\prime}\sum_{\tau^\prime}\langle ij \sigma\sigma^\prime \tau\tau^\prime | \hat{v}_{ij} |ij \sigma\sigma^\prime \tau\tau^\prime\rangle+U_R,\\
    \end{aligned}
\end{equation}
where $\tau=1$ for neutron and $\tau=-1$ for proton. The summation of $j$ is over all the states within the Fermi space, and $U_R$ is the rearrangement term. The analytical expressions for $U_\tau(k)$ with the expression for the QG interaction as,
\begin{equation}\label{eq:Utau}
   \begin{aligned}
       U_{\tau}& = \sum_{l=1}^{2} \left(\tilde{A}_l+\frac{1}{2}\tilde{B}_l\right) (\sqrt{\pi}\mu_l)^3 \rho+\tau\sum_{l=1}^{2} \frac{1}{2}\tilde{B}_l (\sqrt{\pi}\mu_l)^3 \rho\delta\\
       &+2 \sum_{l=1}^{2} \frac{\tilde{C}_l}{\sqrt{\pi}}\bigg[{g}(k,\mu_l,k_{f\tau})+{g}(k,\mu_l,k_{f-\tau})\bigg]\\
       &+2 \sum_{l=1}^{2} \frac{\tilde{D}_l}{\sqrt{\pi}}{g}(k,\mu_l,k_{f\tau})+\left(\beta_0+\frac{\beta_1}{2}\right)\rho^{\sigma+1}\\        
       &+\tau\frac{\beta_1}{2}\rho^{\sigma+1}\delta+\frac{1}{4}\sigma\rho^{\sigma+1}\left[2\beta_0+\beta_1(1+\delta^2)\right].
   \end{aligned}
\end{equation}
Here, $k_{f\tau}=(3\pi^2\rho_\tau)^{1/3}$ and the expression of ${g}(k,\mu_l,k_{f\tau})$ is,
\begin{equation}
\begin{aligned}
        &{g}(k,\mu_l,k_{f\tau}) =\\
        &\quad \frac{1}{\mu_l k}\bigg(e^{-\frac{\mu_l^2(k_{f\tau}+k)^2}{4}}-e^{-\frac{\mu_l^2(k_{f\tau}-k)^2}{4}}\bigg) \\
    &+ \frac{\sqrt{\pi}}{2}\bigg[\mathrm{erf}\left(\frac{\mu_l}{2}(k_{f\tau}-k)\right)+\mathrm{erf}\left(\frac{\mu_l}{2}(k_{f\tau}+k)\right)\bigg].
\end{aligned}
\end{equation}
The shape of $g(k,\mu_l,k_{f\tau})$ is like a Gaussian. The first and second terms in Eq.(\ref{eq:Utau}) originate from the direct term for the finite two-body interaction, and the third and fourth terms are caused by the exchange terms for two-body interaction. The fifth and the sixth term originates from the non-linear density-dependent term, and the last term in Eq.(\ref{eq:Utau}) is the rearrangement term $U_R$. 

\begin{figure}[htbp]
    \centering
    \includegraphics[width=1\linewidth]{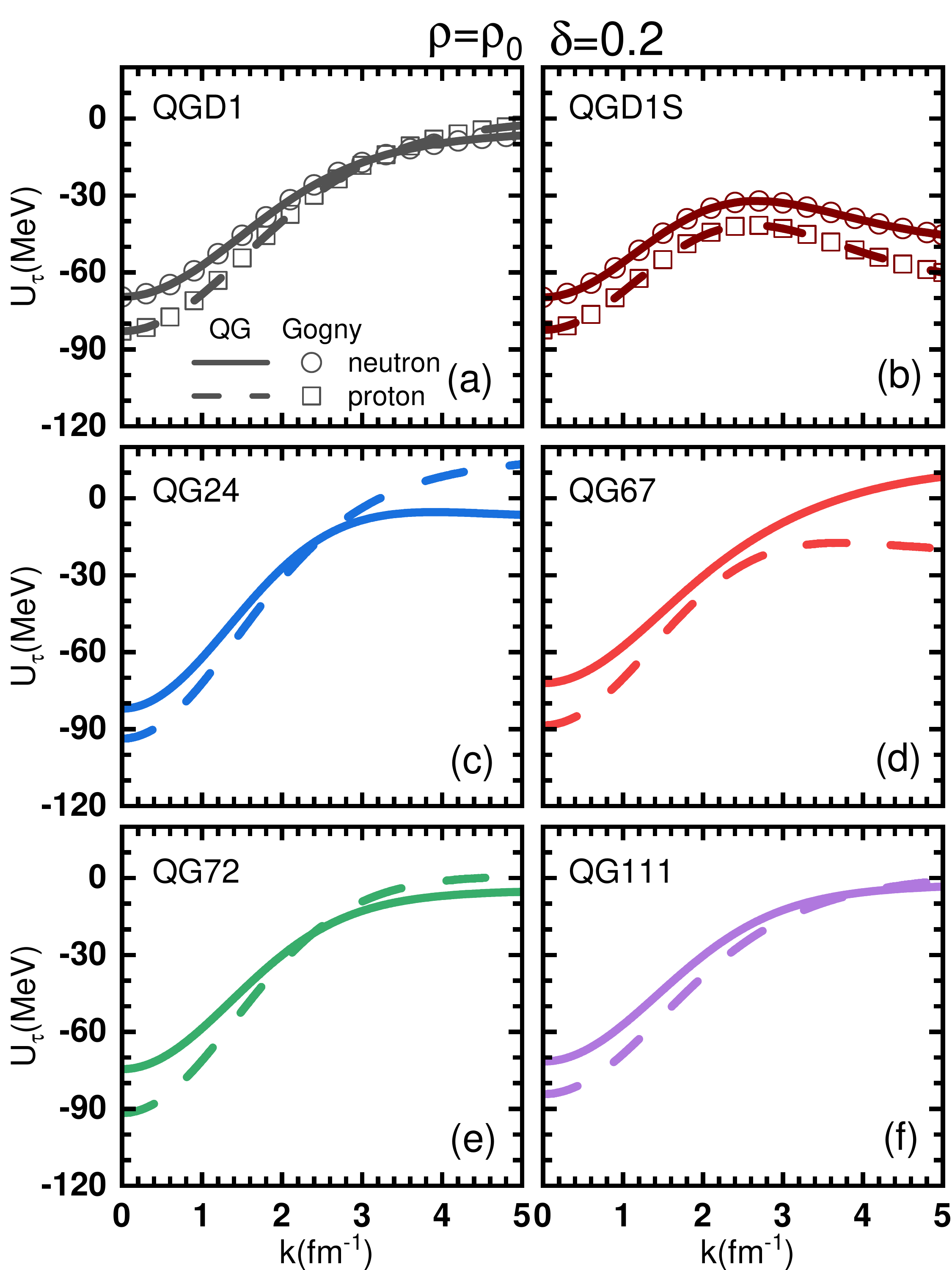}
    \caption{(Color online) Single-particle potential as a function of momentum for neutrons and protons. The results are obtained at $\rho=\rho_0$ and isospin asymmetry $\delta=0.2$. Different panels are for different QG interaction parameter sets. 
    }
    \label{fig:U_tau}
\end{figure}

In Fig. \ref{fig:U_tau}, we present the $U_n$ and $U_p$ in the nuclear matter at normal density $\rho_0$ and isospin asymmetry $\delta=0.2$. The different panels correspond to different parameter sets. Panel (a) and (b) show the $U_n$ and $U_p$ for QGD1 and QGD1S, which are consistent with the results of Gogny D1 and D1S as in Ref.\cite{Roshan2014PRC}. 
Below $k\approx 2-3 \mathrm{fm}^{-1}$, all the $U_\tau$ increases with the momentum for all the parameter sets we used. Above $k$=2-3 $\mathrm{fm}^{-1}$, the situation becomes complicated. For example, $U_n$ start to decrease with momentum for QGD1S and QG24, while $U_p$ decreases with momentum for QGD1S and QG67. These behaviors are mainly attributed to the different widths of $g(k,\mu_1, k_{f\tau})$ and $g(k,\mu_2, k_{f\tau})$, as well as the values of $\tilde{C}_l$ and $\tilde{D}_l$.

\begin{figure}[htbp]
    \centering
    \includegraphics[width=1\linewidth]{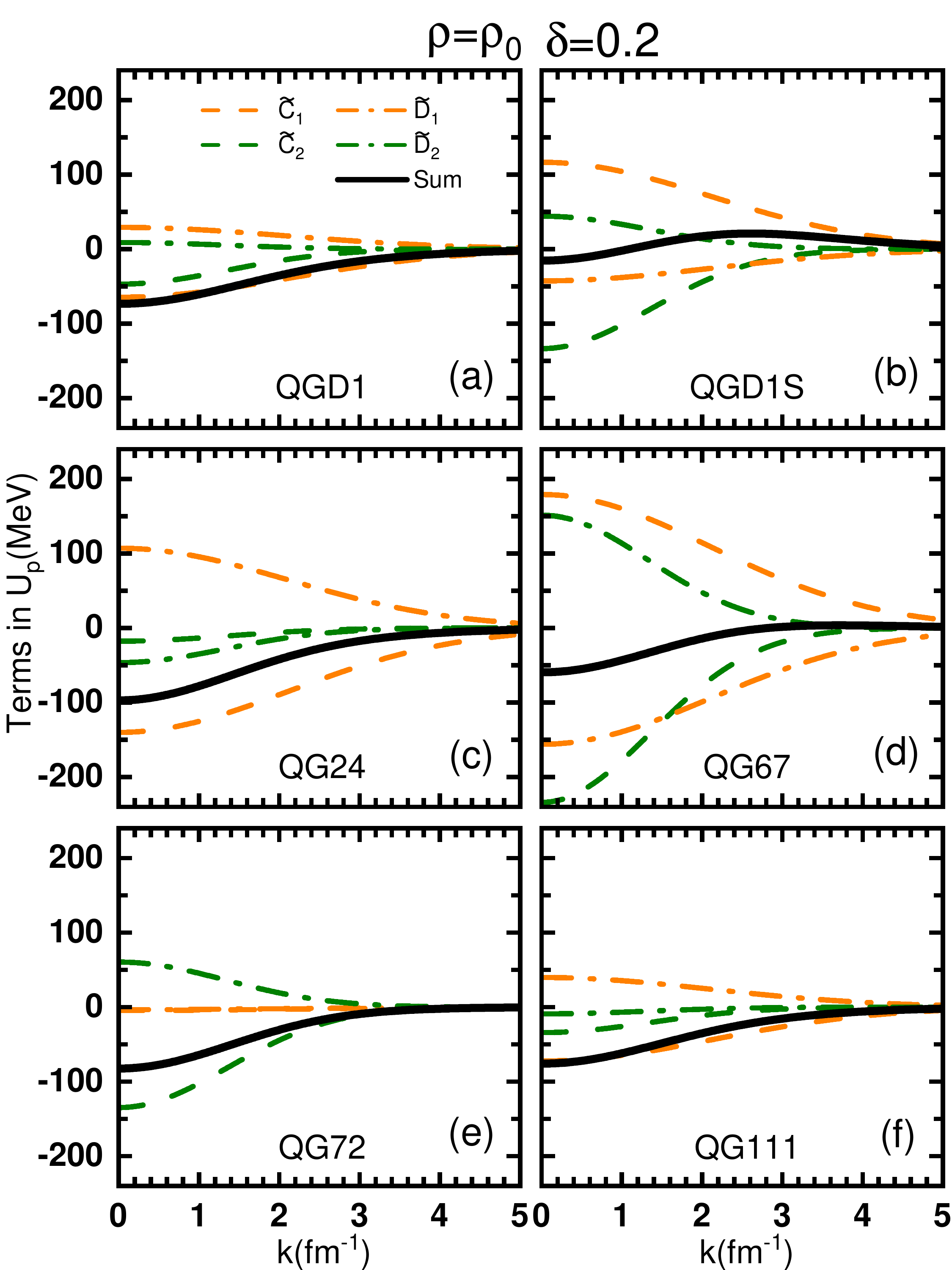}
    \caption{(Color online) Momentum dependent terms in the formula of $U_p$. The results are obtained at $\rho=\rho_0$ and $\delta=0.2$. Different panels are for different parameter sets. }
    \label{fig:CK&DK}
\end{figure}

To quantitatively understand how the $U_p$ evolves with momentum for different QG interactions, we plot the terms related to $\tilde{C}_1$, $\tilde{C}_2$, $\tilde{D}_1$, and $\tilde{D}_2$ in Eq.(\ref{eq:Utau}) in Fig.~\ref{fig:CK&DK} (a)-(f). The black curves are the sum of the $\tilde{C}_1$, $\tilde{C}_2$, $\tilde{D}_1$, and $\tilde{D}_2$ terms, each depicted in different colors. Clearly, when the absolute values of the $\tilde{C}_1$ and $\tilde{D}_1$ terms are much larger (or smaller) than those for $\tilde{C}_2$ and $\tilde{D}_2$ terms, the $U_p$ monotonically increase with momentum. When the crossover occurs between the $\tilde{C}_2$ term and $\tilde{D}_1$ term, as in panel (b) and (d), the $U_p$ exhibits non-monotonic behavior. Similarly, one can expect the behavior of $U_n$ for certain values of $\tilde{C}_1$, $\tilde{C}_2$, $\tilde{D}_1$, and $\tilde{D}_2$.


\begin{figure}[htbp]
    \centering
    \includegraphics[width=1\linewidth]{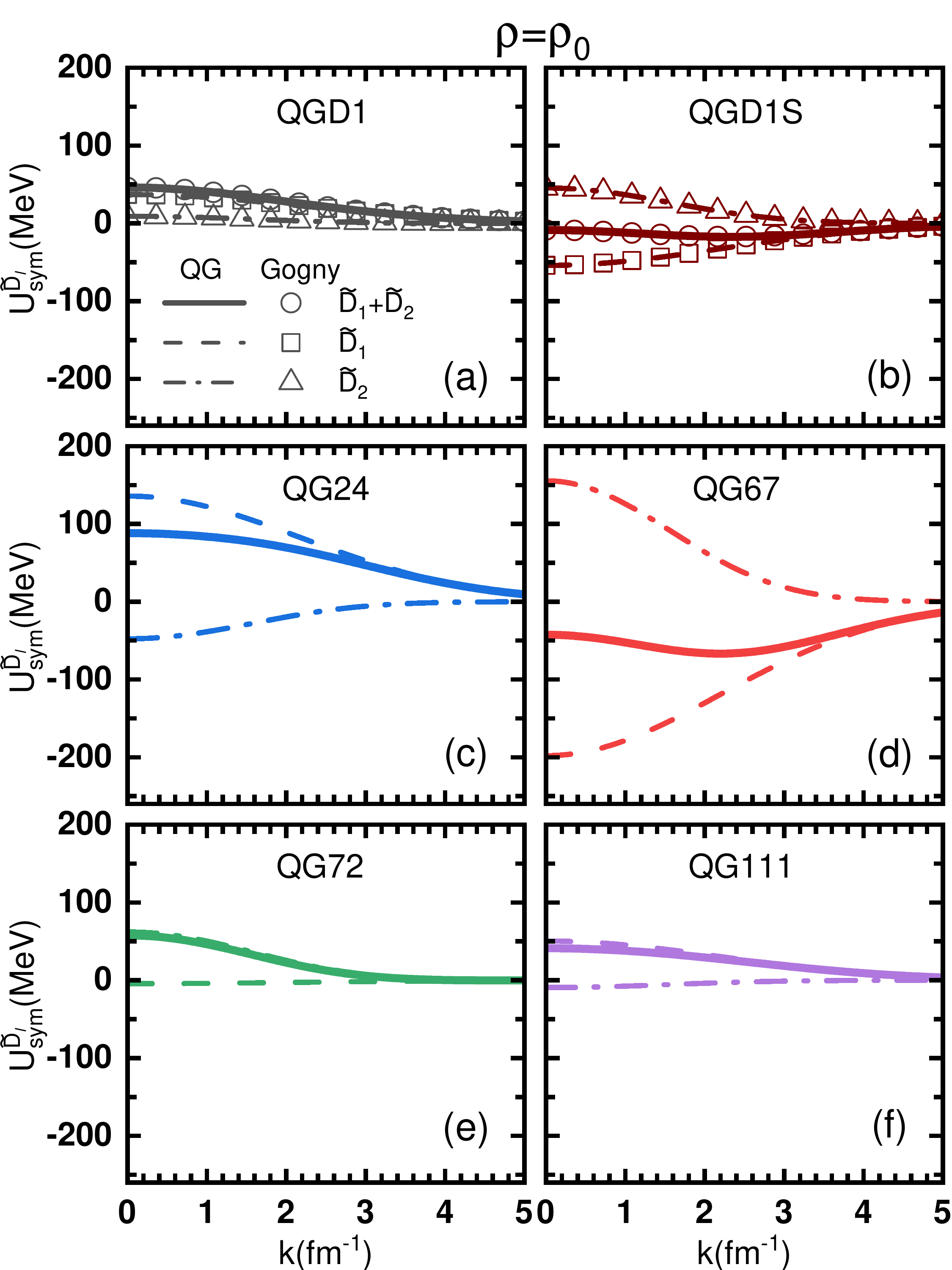}
    \caption{(Color online) Momentum dependent term in the expression of symmetry potential, i.e., $U^{\tilde{D}_1}_{sym}$, $U^{\tilde{D}_2}_{sym}$, and $U^{\tilde{D}_1}_{sym}+U^{\tilde{D}_2}_{sym}$ for different parameter sets. The results are obtained at $\rho=\rho_0$ and $\delta=0.2$. Different panels are for different parameter sets.}
    \label{fig:U_sym_D}
\end{figure}

Further, we investigate the momentum-dependent symmetry potential $U_{sym}$ which is obtained from $U_n$ and $U_p$ as,
\begin{equation}\label{eq:Usym}
    \begin{aligned}
        U_{sym}
        & = \frac{U_n-U_p}{2\delta}\\
        & = \sum_{l=1}^2\frac{\tilde{B}_l}{2}(\sqrt{\pi}\mu_l)^3\rho+\frac{\beta_1}{2}\rho^{\sigma+1}\\
        &\quad+\frac{1}{\delta}\sum_{l=1}^2\frac{\tilde{D}_l}{\sqrt{\pi}}\left[{g}(\mu_l,k,k_{fn})-{g}(\mu_l,k,k_{fp})\right].
    \end{aligned}
\end{equation}
Since the first and second terms in Eq.(\ref{eq:Usym}) are independent of the momentum, we only plot the third term of $U^{\tilde{D}_l}_{sym}$ in Fig. \ref{fig:U_sym_D}. Similar to Fig.~\ref{fig:U_tau}, different panels are for different QG interaction parameter sets. The momentum-dependent symmetry potential $U_{sym}$ obtained with QG67 and D1S has a shape of first decreasing and increasing thereafter. It can be attributed to the crossover between the terms of $\tilde{C}_2$ and $\tilde{D}_1$ as discussed above.

As the nucleon effective mass is determined by the single-particle potential, it is important to examine the effective mass for the QG interaction. The expression of effective mass reads
\begin{equation}
    \begin{aligned}
        \frac{m}{m_\tau^*}&= \left( 1+\frac{m}{p}\frac{\partial U_\tau}{\partial p}\right)|_{\rho_0}\\
        & = 1+\frac{m}{\hbar^2}\bigg\{ \sum_{l=1}^{2} \frac{\tilde{C}_l}{\sqrt{\pi}}\mu_l^2\Bigg[{h}(\mu_lk,\mu_lk_{fn})+{h}(\mu_lk,\mu_lk_{fp})\Bigg]\\
        &\quad + \sum_{l=1}^{2} \frac{\tilde{D}_l}{\sqrt{\pi}} \mu_l^2\left[{h}(\mu_lk,\mu_lk_{f\tau})\right]\bigg\}.
    \end{aligned}
\end{equation}
Here, the function ${h}$ is,
\begin{equation}
\begin{aligned}
         {h}(\eta,\eta_{\tau})
        & =\frac{1}{\eta^3}\bigg\{ (-2-\eta\eta_{\tau})\exp\bigg[{-\frac{(\eta_{\tau}+\eta)^2}{4}}\bigg]\\
        &\quad+(2-\eta\eta_{\tau})\exp\bigg[{-\frac{(\eta_{\tau}-\eta)^2}{4}}\bigg] \bigg\}.
\end{aligned}
\end{equation}
In Table~\ref{tab:parameters}, we list the values of the effective masses $m^*/m$ at normal density and Fermi momentum. The values are in 0.57-0.72 for the sets we studied.

\begin{figure}
    \centering
    \includegraphics[width=1\linewidth]{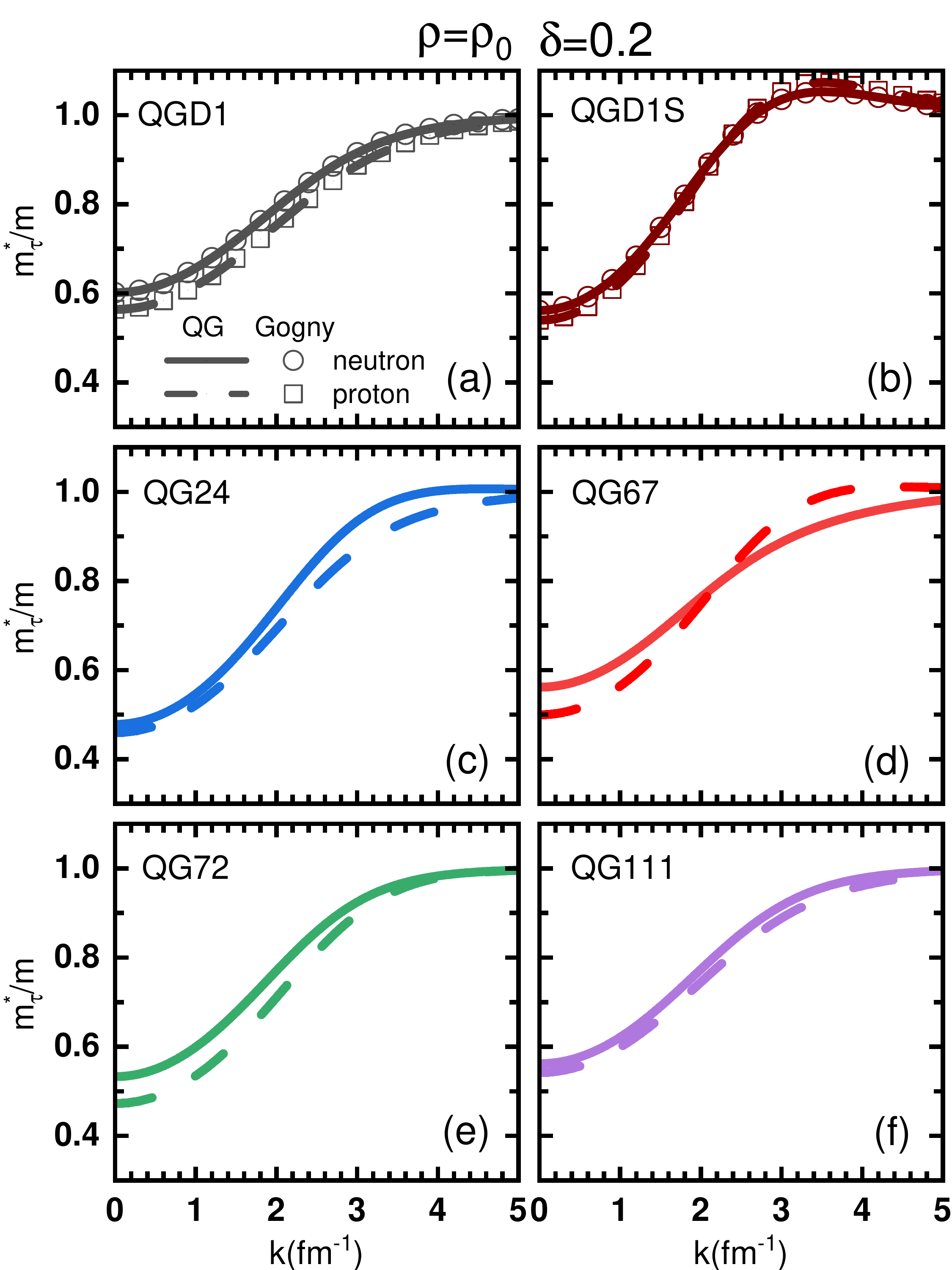}
    \caption{(Color online) The momentum dependence of the effective mass. The results are obtained at $\rho=\rho_0$ and $\delta=0.2$. Different panels are for different parameter sets.}
    \label{fig:E_mass}
\end{figure}

To understand the momentum dependence of the effective mass for neutrons and protons, we present them at normal density in Fig. \ref{fig:E_mass}. Generally, the $m_\tau^{*}/m$ increases with momentum when $k<3$ fm$^{-1}$. In addition, all the parameter sets predict the $m_n^*>m_p^*$ over the entire momentum range except for the parameter sets QG67 and QGD1S. As discussed previously, both of these sets predict the non-monotonic momentum-dependent symmetry potential which means that the effective mass splitting will turn over as momentum increases. For both sets, we found that $m_n^*>m_p^*$ at $k<2.0$ fm$^{-1}$ and $m_n^*<m_p^*$ at $k>2.0$ fm$^{-1}$. This behavior differs from what we observe with the zero-range Skyrme interaction, where the momentum dependent symmetry potential is monotonic, but it is similar to the predictions from Relativistic Hartree-Fock calculations\cite{WHLONG2006PLB}. Consequently, it is interesting to explore whether this prediction is true or not. One way to investigate this behavior is to investigate the different forms of the momentum-dependent symmetry potential via heavy-ion collisions (HICs). This naturally requires incorporating the QG interaction into the transport models.

To realize the QG interaction in the practical QMD-like models, we first need the expression of the potential energy with the QMD wave function. It is straightforward to get the potential energy in the QMD-like model, and it reads,
\begin{equation}\label{eq:Uqmd}
\begin{aligned}
&V=\frac{1}{2}\sum_{ij}\langle\Phi|v_{ij}|\Phi \rangle\\
& = 2\sum_{l=1}^{2} \Bigg\{  \left[\tilde{A}_l \sum_{i,j=1,j\ne i}^{A}  + \tilde{B}_l\left(\sum_{i,j\in n, i\ne j}+\sum_{i,j\in p, i\ne j}\right) \right]\\
&\quad\exp\bigg[{-\frac{(\mathbf{r}_{i}-\mathbf{r}_j)^2}{4\sigma_r^2+\mu_l^2}}\bigg](\frac{\mu_l^2}{4\sigma_r^2+\mu_l^2})^{\frac{3}{2}} \\
&\quad+  \left[\tilde{C}_l\sum_{i,j=1,j\ne i}^{A}  + \tilde{D}_l\left(\sum_{i,j\in n, i\ne j}+\sum_{i,j\in p, i\ne j}\right) \right] \\
&\quad\exp\bigg[{-\frac{(\mathbf{r}_{i}-\mathbf{r}_j)^2}{4\sigma_r^2+\mu_l^2}}\bigg](\frac{\mu_l^2}{4\sigma_r^2+\mu_l^2})^{\frac{3}{2}}\exp\left[{-\frac{\mu_l^2(\mathbf{k_i}-\mathbf{k}_j)^2}{4}}\right] \Bigg\}\\
&\quad+2\beta_0\int \rho^2(\mathbf{r})\rho^{\sigma}  (\mathbf{r})d^3 \mathbf{r}\\
&\quad+2\beta_1\left[\int \rho^2_n(\mathbf{r})\rho^{\sigma}  (\mathbf{r})d^3 \mathbf{r}+ \int \rho^2_p(\mathbf{r})\rho^{\sigma}  (\mathbf{r})d^3 \mathbf{r}\right].
\end{aligned}
\end{equation}
Here, the density $\rho$ and $\rho_{n/p}$ are,
\begin{equation}
   \begin{aligned}
    &\rho(\mathbf{r})
   	 = \sum_{i=1}^A\frac{1}{(2\pi\sigma_r^2)^{\frac{3}{2}}}\exp\bigg[{-\frac{(\mathbf{r}-\mathbf{r}_{i})^2}{2\sigma_r^2}}\bigg],\\
    &\rho_{n/p}(\mathbf{r})
   	 = \sum_{i\in n,p}\frac{1}{(2\pi\sigma_r^2)^{\frac{3}{2}}}\exp\bigg[{-\frac{(\mathbf{r}-\mathbf{r}_{i})^2}{2\sigma_r^2}}\bigg].
   \end{aligned}
\end{equation}
Based on the expression of Eq.(\ref{eq:Uqmd}), one can easily derive the equation of motion for nucleons and expect that the stability of the initial nucleus in the QMD-like models could be enhanced\cite{JPYANG2021PRC} owing to the following reason. The width of the Gaussian functions in Eq.(\ref{eq:Uqmd}) is $4\sigma_r^2+\mu_l^2$ with the QG interaction, which is larger than the width of the Gaussian function $4\sigma_r^2$ in the model with the Skyrme interaction\cite{YXZhang2020FOP}. The inclusion of these finite-range interactions in the improved quantum molecular dynamics model (ImQMD) is currently in progress.


In summary, we propose an effective finite-range interaction for quantum molecular dynamics (QMD)-like models, i.e., QG interaction. This interaction can be directly derived from the conventional Gogny interaction, enabling a deeper understanding or better constraints on the Gogny interaction through heavy-ion collisions. In addition, by fitting the isospin symmetric EOS obtained with QG interaction to the isospin\textcolor{blue}{-}symmetric EOS calculated by the Gogny D1 parameter, we also obtain 13 parameter sets that have different stiffnesses of symmetry energy, momentum-dependent symmetry potential, and effective mass splitting. Impressively, the symmetry potential obtained by utilizing the QG67 parameter set exhibits a non-monotonic momentum-dependent symmetry potential, which is similar to the results obtained with RHF in Ref.\cite{WHLONG2006PLB}.

One way to understand this non-monotonic momentum-dependent symmetry potential is to measure isospin-sensitive observables at intermediate energies of heavy-ion collisions and compare them with transport model calculations. 
Incorporating this finite-range interaction into the improved quantum molecular dynamics model (ImQMD) and analyzing its effects on isospin-sensitive observables are in progress.

\section*{Acknowledgments}

The authors thank the helpful discussions with Prof. Lie-Wen Chen. This work was partly inspired by the transport model evaluation project, and it was supported by the National Natural Science Foundation of China under Grants No. 12275359, No. 12375129, No. 11875323 and No. 11961141003, by the National Key R\&D Program of China under Grant No. 2023 YFA1606402, by the Continuous Basic Scientific Research Project, by funding of the China Institute of Atomic Energy under Grant No. YZ222407001301, No. YZ232604001601, and by the Leading Innovation Project of the CNNC under Grants No. LC192209000701 and No. LC202309000201. We acknowledge support by the computing server SCATP in China Institute of Atomic Energy and Basic Research Special Zone.

\begin{widetext}
\appendix

\section{15 QG interaction parameter sets}
\label{app:15qg}

The 15 QG interaction parameter sets and the corresponding nuclear matter parameters are listed in Table~\ref{tab:parameters} and Table~\ref{tab:slope}. 

\begin{table*}
\caption{\label{tab:parameters}Parameter sets of QMD-Gogny interaction. The dimension of $\tilde{A}_l$ and $\tilde{B}_l$ are in MeV, $\beta_0$ and $\beta_1$ are in MeVfm$^{3\sigma+1}$. }
\begin{ruledtabular}
\begin{tabular}{ccccccccccccc}
      Para. & $\tilde{A}_1$ & $\tilde{A}_2$ & $\tilde{B}_1$ & $\tilde{B}_2$ & $\tilde{C}_1$ & $\tilde{C}_2$& $\tilde{D}_1$ & $\tilde{D}_2$& $\beta_0$ & $\beta_1$  & $\sigma$ & $\chi^2_r$ \\ 
    \hline
        QGD1 & -452.40 & -27.19  & 507.98   & -2.87   & -271.66 & -50.18  & 301.20   & 22.42   & 2025.00 & -2025.00 & 0.33 \\
        QGD1S&-1070.30 & 21.90   & 1114.73  & -50.85  & 490.84  & -142.53 & -439.90  & 111.66  & 2085.90 & -2085.90 &  0.33\\
      \hline
        QG2  & -301.19 & 60.62   & -572.51  & 4.04    & -116.66 & -234.60 & 80.46    & 360.21  & 844.43  & 43.71    & 0.39 & 0.163   \\
        QG7  & 779.82  & -143.00 & 519.63   & -173.18 & -96.84  & -213.43 & 30.58    & 273.85  & 773.83  & -313.95  & 0.50 & 0.953   \\
        QG21 & -860.38 & 110.43  & -663.68  & 166.81  & -294.40 & -64.78  & 571.10   & -18.15  & 1587.92 & -1706.73 & 0.46 & 0.655   \\
        QG22 & 682.08  & -163.61 & 570.36   & -192.62 & 32.90   & -208.61 & -337.47  & 343.37  & 793.73  & 455.14   & 0.33 & 0.005   \\
        QG24 & -293.93 & 29.14   & -315.55  & 57.36   & -589.87 & -18.86  & 1104.65  & -117.22 & 1504.68 & -1685.91 & 0.49 & 0.912   \\
        QG25 & -414.98 & 19.39   & 425.44   & -14.07  & -351.09 & -30.16  & 644.59   & -104.57 & 1755.57 & -2262.27 & 0.52 & 0.966   \\
        QG27 & -532.00 & -8.23   & 385.4    & -26.07  & -147.74 & -104.10 & -187.40  & 182.00  & 1742.54 & -942.12  & 0.25 & 0.560   \\
        QG28 & 661.69  & -168.05 & 657.78   & -168.34 & -45.24  & -162.66 & -1.67    & 208.87  & 956.54  & -190.02  & 0.40 & 0.209   \\
        QG43 & 677.29  & -223.41 & 907.53   & -119.32 & -409.22 & -17.29  & 597.28   & -57.90  & 1691.66 & -1531.71 & 0.36 & 0.071   \\
        QG67 &-1037.68 & 98.49   & -2296.79 & 563.21  & 754.44  & -249.79 & -1609.99 & 382.08  & 835.08  & 28.93    & 0.40 & 0.228   \\
        QG72 &-345.80  & 50.35   & -436.17  & 34.03   & -19.10  & -143.92 & -35.67   & 152.53  & 407.00  & 715.99   & 0.44 & 0.452   \\
        QG75 &-307.26  & 38.07   & -509.35  & 19.94   & -81.11  & -134.56 & -92.85   & 187.69  & 432.76  & 1091.01  & 0.34 & 0.020   \\
        QG111&-229.07  & 29.63   & -245.61  & -36.48  & -304.68 & -36.51  & 410.17   & -22.90  & 231.28  & 1373.68  & 0.36 & 0.073  \\
\end{tabular}
\end{ruledtabular}
\end{table*}

\begin{table*}
\caption{\label{tab:slope} Properties of nuclear matter as predicted by QMD-Gogny interaction.}
\begin{ruledtabular}
\begin{tabular}{cccccccccccc}
No. & Para. & $\rho_0$ & $\frac{E_0}{A}$ & $K_0$&$S_0$ &$L$ &$\frac{m_\tau^*}{m}$ & $m_n^*$ &$m_p^*$ &$\frac{m_n^*-m_p^*}{\delta}$& \\
 & & (fm$^{-3}$)&(MeV/u)& (MeV) & (MeV) & (MeV) &  ($\delta=0$) & ($\delta=0.2$) & ($\delta=0.2$) & ($\delta=0.2$) & \\  \hline
1    & QGD1  & 0.166 & -16.26 & 228.89 & 30.667 & 18.39 & 0.68 & 0.70 & 0.66 & 0.20   \\
2    & QGD1S & 0.163 & -15.96 & 202.45 & 31.100 & 22.46 & 0.70 & 0.72 & 0.70 & 0.10   \\
3    & QG2   & 0.166 & -16.30 & 229.93 & 32.063 & 2.05  & 0.64 & 0.75 & 0.56 & 0.93  \\
4    & QG7   & 0.166 & -16.32 & 231.40 & 29.365 & 6.50  & 0.57 & 0.63 & 0.52 & 0.54  \\
5    & QG21  & 0.166 & -16.31 & 229.76 & 31.540 & 20.89 & 0.61 & 0.63 & 0.59 & 0.18  \\
6    & QG22  & 0.166 & -16.28 & 229.18 & 28.400 & 21.84 & 0.68 & 0.77 & 0.61 & 0.78  \\
7    & QG24  & 0.166 & -16.32 & 230.48 & 32.096 & 24.13 & 0.59 & 0.60 & 0.57 & 0.15  \\
8    & QG25  & 0.166 & -16.32 & 230.33 & 32.274 & 25.25 & 0.58 & 0.58 & 0.58 & 0.02  \\
9    & QG27  & 0.166 & -16.25 & 227.72 & 29.015 & 26.53 & 0.72 & 0.76 & 0.67 & 0.45  \\
10   & QG28  & 0.166 & -16.30 & 229.61 & 30.270 & 27.79 & 0.64 & 0.70 & 0.60 & 0.51  \\
11  & QG43  & 0.166 & -16.29 & 229.84 & 33.552 & 43.27 & 0.66 & 0.67 & 0.65 & 0.12  \\
12   & QG67  & 0.166 & -16.30 & 229.74 & 32.908 & 66.51 & 0.64 & 0.67 & 0.62 & 0.24  \\
13   & QG72  & 0.166 & -16.31 & 229.99 & 32.884 & 71.96 & 0.62 & 0.65 & 0.59 & 0.33  \\
14   & QG75  & 0.166 & -16.30 & 229.50 & 32.968 & 75.44 & 0.67 & 0.72 & 0.63 & 0.45  \\
15   & QG111 & 0.166 & -16.29 & 229.84 & 33.412 & 110.84& 0.65 & 0.67 & 0.64 & 0.12 \\
\end{tabular}
\end{ruledtabular}
\end{table*}
\end{widetext}

\bibliography{apssamp}

\end{document}